\documentclass[aps,prep,epsfile,nofootinbib]{revtex4}

\usepackage{graphics}
\usepackage{slashed}
\usepackage{amssymb}
\usepackage{amsmath}
\usepackage{graphicx}
\begin{document}

\title{Quantum Gravity and Inflation as an Open System: A New Paradigm}
\author{$^{1}$ Alan Sebasti\'an Morales\footnote{alanmorales@mdp.edu.ar}, $^{1,2}$ Mauricio Bellini
\footnote{mbellini@mdp.edu.ar},$^{1,2}$ Juan Ignacio Musmarra\footnote{jmusmarra@mdp.edu.ar}}
\address{
$^1$ Instituto de Investigaciones F\'{\i}sicas de Mar del Plata (IFIMAR), \\
Consejo Nacional de Investigaciones Cient\'ificas y T\'ecnicas (CONICET), Mar del Plata, Argentina.\\
$^2$ Departamento de F\'{\i}sica, Facultad de Ciencias Exactas y
Naturales, Universidad Nacional de Mar del Plata, Funes 3350, C.P. 7600, Mar del Plata, Argentina.}
\begin{abstract}
As part of our program to develop a general theory of relativity for open systems, we introduce a covariant theory that incorporates the effects of classical and quantum spacetime alterations in a new metric tensor that effectively includes these alterations, thereby generating a Riemannian manifold from a new varied action without boundary terms. We illustrate the theory by studying an inflationary model which incorporates the quantum feedback effects of spacetime on the dynamics of the inflaton field $\hat{\varphi}$, enabling the simultaneous quantization of $\hat{\varphi}$ and the fluctuating gravitational field $\hat{\Omega}$ without relying on perturbative theory. We obtain an exact solution for
the modes of both fields, $\hat{\varphi}$ and $\hat{\Omega}$, which comply with different quantum algebras. The normalization of the inflaton field modes that is obtained is intrinsically related to the geometric field modes, in such a way that the quantization of the fields results from an expression that links the geometric fields with the physical fields. Finally, the quadratic fluctuations of spacetime are calculated, and its spectrum is analyzed.
\end{abstract}
\maketitle

\section{Introduction, motivation and preliminaries}

Cosmic inflation \cite{in0,in1,in11,in2,in3,in33,in333,hm,in4,b1,bcms,vega1,maube1,maube2,vega2,vega3,vega4,mrv} explains a period of exceedingly rapid exponential expansion in the early universe, effectively resolving issues like the flatness and horizon problems. However, this theory notably lacks a fundamental explanation for both the onset and cessation of this crucial phase. In this context, several efforts have been devoted to identifying the viable initial conditions that could give rise to the inflationary expansion\cite{Kundu2011,Salvio2017,Cai2019,Melcher2023,Joana2024}, within both extensions of General Relativity\cite{Cognola2009,Odintsov2021} and candidate theories of quantum gravity\cite{Powell2006,Wang2007,Destri2009,Dudas2012,Das2014,Kitazawa2014,Gruppuso2016,Jin2018}. Potential insights into this issue may be gained from observations of the cosmic microwave background\cite{Agullo2013,Ashtekar2015,Zhu2017,Mohammadi2024,Shahalam2025}. Nevertheless, it is this fundamental gap that underscores the necessity of quantum gravity in the cosmological models \cite{boj,cher,bit,day}. Therefore, contemporary cosmic inflation theories are increasingly incorporating quantum gravity effects, not only to overcome the aforementioned limitations of the standard model but also to offer a more complete description of the very early universe, a regime where extreme conditions unequivocally demanded a foundational role for both gravitational and quantum mechanical principles in a relativistic framework. If we could find an inflationary model consistent with observations and derived from a theory of quantum gravity, we would move closer to a unified understanding of the origin and evolution of the universe, connecting particle physics, cosmology, and quantum gravity \cite{el}.

A significant challenge in modern cosmology is to achieve a non-perturbative description of the dynamics of both geometric and physical quantum fields. While Einstein's equations provide a classical description of a system's dynamics on a Riemannian manifold, we must consider both spacetime and the physical fields that describe the system as quantum in origin. Their expectation values on the classical relativistic background are compatible with General Relativity, but we must take into account that quantum fluctuations relative to their expectation values on the Riemannian background will impact the system's global (or classical) dynamics. This effect can be treated as a flow of a geometrical field through a closed hypersurface when the action is varied, appearing as boundary terms. To treat this flow mathematically, we can expand the Riemannian manifold using a recently introduced covariant derivative \cite{Bellini}. For a non-perturbative theory, these boundary terms must be included in the varied action, which  must be preserved. On the other hand, since relativistic spacetime describes a finite region of causally connected events that dynamically changes under quantum fluctuations, the causally connected region of the universe must be considered an open system altered by its environment. This environment consists of the fluctuations accounted for in the boundary terms of the varied action. Although the action will be treated from its quantum nature, its integrand must be constructed from integrable (or classical) physical and geometric quantities.

This work begins with a recently proposed extension of the general theory of relativity. This extension incorporates an additional term in Einstein's effective equations on the Riemann manifold, derived from the flow of a geometric quantum field within the varied Einstein-Hilbert (EH) action. It is a standard practice to include counterterms in the action to cancel the contribution of the last term in the varied action, thereby ensuring a well-defined variational principle \cite{York, Gibbons, Parattu}. However, there is another possibility that can be useful to study open systems in GR which was developed in \cite{rb1}, and other subsequent works. This new term introduces a cosmological function and produces an effective alteration of the background metric tensor \cite{univ}. From this new metric, a novel relativistic theory can be constructed that accounts for the geometric fluctuations of spacetime on the background. We will then use this new background spacetime to describe the dynamics of the inflaton field, which can subsequently be quantized by considering the quantum fluctuations of spacetime. The manuscript is organised as follows: in Sect. II we refresh the formalization of General relativity in the case where the boundary terms are included in the EH varied action. In Sect. III we introduce a new General Relativity (GR) formalism without boundary terms, such that spacetime fluctuations are included in the line element as a background geometry. In Sect. IV we illustrate this formalism by developing an inflationary model where the expansion of the universe is driven by a massive scalar field (inflaton) and we study its dynamics on the new Riemannian manifold where the tensor metric is $\bar{g}_{\alpha\beta}$. We shall propose a particular model for the inflation of the universe where we will find exact solutions for the fields $\hat{\varphi}$ and $\hat{\Omega}$. We shall quantize the field taking into account quantum gravity effects due to the inflaton field without using the semiclassical approximation, to later calculate the mean square fluctuations associated with the gravitational field $\hat{\Omega}$ and analyze its spectrum. Finally, in Sect. V we develop some final comments.

\section{General Relativity with boundary terms included}

We shall consider that the background expectation value of the Lagrangian density on the Riemann manifold: $\left<B\right|\hat{\cal L}_m\left|B\right>$ is calculated by using the quantum state $\left|B\right>$, which is static, because we are dealing with the Heisenberg representation, so that the quantum operators evolve in spacetime. Furthermore, $R=g^{\alpha\beta}\,R_{\alpha\beta}$ is the scalar curvature, $R_{\alpha\beta}$ is the Ricci tensor, $g_{\alpha\beta}$ is the covariant metric tensor with determinant $g$, the signature is $(+,-,-,-)$, and $\nabla_{\mu}g_{\alpha\beta}=0$ on the Riemann manifold, in which $g_{\alpha\beta}$ is the fundamental tensor. We shall consider the EH action. Since the Lagrangian density will be written in quantum fields, we must consider in action its expectation value on the Riemannian manifold, since it is constructed from classical quantities:
\begin{equation}\label{action}
{\cal I} = \int\,d^4x\,\sqrt{-g}\,\left\{\frac{R}{2\kappa}+\left<B\right|\hat{\cal L}_m\left|B\right>\right\}.
\end{equation}
This means that the covariant derivatives on this manifold will be defined with respect to the Levi-Civita connections:
\begin{tiny}$\left\{ \begin{array}{cc}  \alpha \, \\ \beta \, \nu  \end{array} \right\}$\end{tiny} \footnote{
\begin{displaymath}
\left\{\begin{array}{cc}  \alpha \, \\ \beta \, \nu  \end{array} \right\}=\frac{1}{2} g^{\alpha\mu} \left[g_{\mu\beta,\nu} + g_{\nu\mu,\beta} - g_{\beta\nu,\mu}\right].
\end{displaymath}
We shall consider the Ricci tensor defined as
\begin{displaymath}
R_{\alpha\beta} \equiv R^{\nu}_{\alpha\beta\nu}= \left\{ \begin{array}{cc}  \nu \, \\ \alpha \, \nu  \end{array} \right\}_{,\beta}-\left\{ \begin{array}{cc}  \nu \, \\ \alpha \, \beta  \end{array} \right\}_{,\nu} + \left\{ \begin{array}{cc}  \nu \, \\ \beta \, \mu  \end{array} \right\} \left\{ \begin{array}{cc}  \mu \, \\ \nu \, \alpha  \end{array} \right\} -
\left\{ \begin{array}{cc}  \mu \, \\ \mu \, \nu  \end{array} \right\} \left\{ \begin{array}{cc}  \nu \, \\ \alpha \, \beta  \end{array} \right\} ,
\end{displaymath}
where $(...)_{,\beta}$ denotes the partial derivative of $(...)$, with respect to $x^{\beta}$.}. The Einstein equations on the background metric are obtained after varying the EH action (\ref{action})
\begin{equation}\label{delta0}
\delta {\cal I} = \int d^4 x \sqrt{-g} \left[{\delta g}^{\alpha\beta}\,G_{\alpha\beta}+  \left<B\right|{\delta g}^{\alpha\beta}  \kappa\, \hat{T}_{\alpha\beta}+ \,g^{\alpha\beta} \hat{\delta R}_{\alpha\beta}\left|B\right> \right]=0,
\end{equation}
where the last terms in (\ref{delta0}) are the boundary terms, which will be considered of quantum nature: \begin{equation}\label{RR}
\delta\Theta\equiv\left<B\right|\hat{\delta\Theta}\left|B\right>=\left<B\right|g^{\alpha\beta} \hat{\delta R}_{\alpha\beta}\left|B\right>.
\end{equation}
To integrate boundary terms into the global dynamics we shall consider that variations in the metric tensor
$\hat{\delta g}_{\alpha\beta}$ are the sole geometric source of curvature variations $\hat{\delta R}_{\alpha\beta}$. For this assumption to hold within a relativistic framework, we shall propose a general relativistic Ricci flow in which $\hat{\delta g}_{\alpha\beta}$ is originated by the geometric alterations of the Riemann manifold caused by a geometric quantum scalar field $\hat{\Omega}$ \cite{ricci1,ricci-flow} \footnote{To describe the quantum flux in (\ref{RR}), we shall consider the varied Ricci quantum tensor, which we will define as an extension of the Palatini expression \cite{pal}:
\begin{equation}\label{ricci}
\hat{\delta R}_{\alpha\beta} = b\,\left[\left(\hat{ \delta\Gamma}^{\mu}_{\alpha\mu} \right)_{\| \beta} - \left(\hat{ \delta\Gamma}^{\mu}_{\alpha\beta}\right)_{\| \mu}\right],
\end{equation}
where $\left( ...\right)_{\| \beta}$ denotes de covariant derivative of $\left( ...\right)$ on the extended manifold \cite{Bellini}.
}:
\begin{equation}\label{fff}
\hat{\delta R}_{\alpha\beta}= \lambda\left(x\right)\,\hat{\delta g}_{\alpha\beta},
\end{equation}
such that, if we require that $g^{\alpha\beta}\,\hat{\delta g}_{\alpha\beta}+\hat{\delta g}^{\alpha\beta}\,g_{\alpha\beta}=0$, we obtain that
\begin{equation}\label{fff2}
g^{\alpha\beta}\,\hat{\delta R}_{\alpha\beta}= \lambda\left(x\right)\,g^{\alpha\beta}\,\hat{\delta g}_{\alpha\beta}= -\lambda\left(x\right)\,\hat{\delta g}^{\alpha\beta}\,g_{\alpha\beta}.
\end{equation}
The constant $b$ must be determined and the components $\delta\Gamma^{\alpha}_{\beta\epsilon}$ are the departure of connections with respect to the Levi-Civita ones. Before introducing the quantum geometric and physical concepts, it is important to define the notions of the covariant derivative on the extended manifold that we will use throughout this work: the covariant derivative of a $(n)$-times contravariant and $(m)$-times covariant mixture tensor $\hat{\Upsilon}^{\alpha_1...\alpha_n}_{\beta_1 ... \beta_m}$, on the extended manifold, the $(n+m+1)$-range quantum tensor:
\begin{eqnarray}
&& \hat{\Upsilon}^{\alpha_1...\alpha_n}_{\beta_1 ... \beta_m \|\mu} = \hat{\nabla}_{\mu} \hat{\Upsilon}^{\alpha_1...\alpha_n}_{\beta_1 ... \beta_m} +
\sum_{i=1}^{n} \hat{\delta\Gamma}^{\alpha_i}_{\nu\mu}\,\hat{\Upsilon}^{\alpha_1..\alpha_{i-1}\nu\alpha_{i+1} ..\alpha_n}_{\beta_1 ... \beta_m}
-\sum_{i=1}^{m} \hat{\delta\Gamma}^{\nu}_{\mu \beta_i}\,\hat{\Upsilon}^{\alpha_1...\alpha_n}_{\beta_1 ..\beta_{i-1}\nu\beta_{i+1}. \beta_m}  \nonumber \\
&& -\eta \sum_{i=1}^{n-1} \left(\hat{\Upsilon}^{\alpha_1..\alpha_i\alpha_{i+1}..\alpha_n}_{\beta_1 ... \beta_m}\,\hat{\Omega}_{\mu}
+\hat{\Omega}_{\mu}\,\hat{\Upsilon}^{\alpha_1..\alpha_{i+1}\alpha_i..\alpha_n}_{\beta_1 ... \beta_m}\right) + \eta \sum_{i=1}^{m-1}\left(\hat{\Upsilon}^{\alpha_1...\alpha_n}_{\beta_1 ..\beta_i\beta_{i+1}.. \beta_m}\, \hat{\Omega}_{\mu}+
\hat{\Omega}_{\mu}\,\hat{\Upsilon}^{\alpha_1...\alpha_n}_{\beta_1 ..\beta_{i+1}\beta_i.. \beta_m}\right). \label{uau}
\end{eqnarray}
The terms in the last row of (\ref{uau}) describe the interaction of the quantum tensor $\hat{\Upsilon}^{\alpha_1...\alpha_n}_{\beta_1 ... \beta_m}$ with the extended manifold. Furthermore, $\eta$ is a parameter to be determined which is related to the quantum effects and $b=z\eta$ is related to the classical ones. When $z\ll 1$ the interaction of the derived field with the extended manifold is quantum dominance. However, when $z \rightarrow \infty$, such interaction is dominated by classical fluctuations. Furthermore, we can define the quantum variation of the quantum tensor field $\hat{\Upsilon}^{\alpha_1...\alpha_n}_{\beta_1 ... \beta_m }$:
\begin{displaymath}
\hat{\delta\Upsilon}^{\alpha_1...\alpha_n}_{\beta_1 ... \beta_m}=\hat{\Upsilon}^{\alpha_1...\alpha_n}_{\beta_1 ... \beta_m \|\mu}\,\bar{U}^{\mu},
\end{displaymath}
where $\bar{U}^{\mu}$ are the components of relativistic velocity, which are solutions of the geodesic equations on the extended manifold: $\delta \bar{U}_{\mu}=0$, or
\begin{equation}\label{geo}
\frac{d \bar{U}^{\alpha}}{d S} + \Gamma^{\alpha}_{\mu\nu} \,\bar{U}^{\mu}\,\bar{U}^{\nu} = \left[ \eta-b\right] \,\bar{U}^{\alpha}\,\left(\bar{U}_{\mu}\bar{U}^{\mu}\right).
\end{equation}
Therefore, we obtain the following expressions for $\hat{\delta g}_{\alpha\beta}$ and $\hat{\delta g}^{\alpha\beta}$:
\begin{equation}
\hat{\delta g}_{\alpha\beta}  = b^{-1}\,\left(\hat{g}_{\alpha\beta}\right)_{\|\mu}\,\bar{U}^{\mu}, \qquad \hat{\delta g}^{\alpha\beta} = b^{-1}\,\left(\hat{g}^{\alpha\beta}\right)_{\|\mu}\,\bar{U}^{\mu},
\end{equation}
such that we can define
\begin{equation}
\bar{g}_{\alpha\beta}  =  {g}_{\alpha\beta}+{\delta g}_{\alpha\beta},\qquad \tilde{g}^{\alpha\beta}  =  {g}^{\alpha\beta}+{\delta g}^{\alpha\beta},
\end{equation}
where ${\delta g}_{\alpha\beta}=\left<B\right|\hat{\delta g}_{\alpha\beta}\left|B\right>$ and ${\delta g}^{\alpha\beta}=\left<B\right|\hat{\delta g}^{\alpha\beta}\left|B\right>$. Notice that $\tilde{g}^{\alpha\beta}$ is not the adjoint of $\bar{g}_{\alpha\beta}$, but ${g}^{\alpha\beta}$ is the adjoint of ${g}_{\alpha\beta}$, such that \cite{p2}
\begin{equation}
g_{\alpha\beta}\,\bar{U}^{\alpha} \bar{U}^{\beta} =\theta(z)-1,
\end{equation}
with
\begin{equation}\label{oro}
\theta(z)= \frac{\left[3z-4+\sqrt{9\,z^2-8\,z}\right]}{4\,(z-1)},
\end{equation}
for $b=z\,\eta$. Furthermore, we shall consider that
\begin{equation}\label{covg}
\bar{g}_{\alpha\beta}\,\bar{U}^{\alpha} \bar{U}^{\beta} =1,
\end{equation}
where
\begin{equation}\label{newmetric}
\bar{g}_{\alpha\beta}=g_{\alpha\beta}\left[1+\frac{2\left[\theta(z)-1\right]}{z} \right] -2\,\bar{U}_{\alpha}\bar{U}_{\beta}.
\end{equation}
From (\ref{RR}) and (\ref{fff}) we obtain the expression for $\lambda(x^{\mu})$
\begin{equation}\label{la}
\lambda(x^{\nu}) = \frac{3z^2\,\eta}{2} \left[\frac{\eta\,(2z+1)\,\left[\bar{\theta}(z)-1\right] + g^{\alpha\beta}\left<B\right|\nabla_{\alpha}\hat{\Omega}_{\beta}\left|B\right>}{ (z-4) \left[\theta(z)-1\right]}\right].
\end{equation}
To make a relativistic quantum description of geometry on the extended manifold, we shall consider the case where
\begin{equation}\label{cone}
\hat{\delta\Gamma}^{\mu}_{\beta\epsilon}=b\,\hat{\Omega}^{\mu}\,g_{\beta\epsilon},
\end{equation}
where $\hat{\Omega}_{\alpha}\equiv \hat{\partial}_{\alpha}\,\hat{\Omega}$ is a quantum operator. The scalar quantum geometric field $\hat{\Omega}$ can be represented by means of a Fourier expansion:
\begin{equation}\label{fu}
\hat{\Omega}(\tau,\vec{r}) = \frac{1}{(2\pi)^{3/2}}\,\int \,d^3 k\,\left[ \hat{\Omega}_k(\tau,\vec{r})+\hat{\Omega}^{\dagger}_k(\tau,\vec{r})\right].
\end{equation}
Here, the modes of the expansion are $\hat{\Omega}_k(\tau,\vec{r})=\hat{A}_k\,{\sigma}(k,\tau,\vec{r})$ and $\hat{\Omega}^{\dagger}_k(\tau,\vec{r})=\hat{A}^{\dagger}_k\,{\sigma}^*(k,\tau,\vec{r})$.  The creation and destruction operators: $\hat{A}^{\dagger}_k$ and $\hat{A}_k$ comply with the algebra:
\begin{equation}\label{m55}
\left<B\left|\left[\hat{A}_k(\tau,\vec{r}),\hat{A}^{\dagger}_{k'}(\tau,\vec{r})\right]\right|B\right>
=\,\delta^{(3)}\left(\vec{k}-\vec{k'}\right),\quad
\left<B\left|\left[\hat{\Omega}^{\dagger}_k(\tau,\vec{r}),\hat{\Omega}^{\dagger}_{k'}(\tau,\vec{r})\right]\right|B\right>=
\left<B\left|\left[\hat{\Omega}_k(\tau,\vec{r}),\hat{\Omega}_{k'}(\tau,\vec{r})\right]\right|B\right>=0.
\end{equation}
Note that the expression on the left in (\ref{m55}) tells us that space-time can be created or annihilated. From a macroscopic point of view, when the distance between two objects increases (decreases), it is due to the creation (annihilation) of space-time between them at a quantum level. As in a previous work \cite{p2}, we shall impose a modified nonlinear quantum algebra \cite{BMAS, mabe}, which is satisfied by the components of $\hat{\Omega}(\tau,\vec{r})$ in the Fourier expansion (\ref{fu}):
\begin{equation}\label{m56}
\left[\frac{\hat{\partial}}{\partial{x}^{\alpha}},\hat{\Omega}_{k'}^\dagger\right]=\,\bar{U}_{\alpha}\,
\left[\hat{\Omega}_{k},\hat{\Omega}_{k'}^\dagger\right].
\end{equation}
The right-hand term in (\ref{m56}) alters the standard geodesic equations when $b \neq \eta$. As we shall see in the next section, this is due to the alterations of space-time produced by its fluctuations that could produce deviations in gravitational lensing or small deviations in planetary orbits.

We introduce this proposal because if we used a linear noncommutative algebra (as usual), the expectation value of $\hat{\delta R}_{\alpha\beta}$ would be zero. This forces us to propose a noncommutative nonlinear algebra that will hold only at the level of geometric quantum fields. Physical fields will obey the usual algebra. The wavenumber components in (\ref{m56}) are given by $k^{\mu}=\left(\frac{mc}{\hbar}\right)\,\bar{U}^{\mu}$ and the altered relativistic velocities with perturbations included, $\bar{U}^{\mu}=\frac{d \bar{x}^{\mu}}{dS}$, are the solutions of the differential equations, $\delta \bar{U}^{\alpha}\equiv \bar{U}^{\beta}\left(\bar{U}^{\alpha}\right)_{\|\beta}=0$, which leaves the $4$-velocity unchanged on the extended manifold:
\begin{equation}\label{geo}
\frac{d \bar{U}^{\alpha}}{d S} + \Gamma^{\alpha}_{\mu\beta} \,\bar{U}^{\mu}\,\bar{U}^{\beta} = \left[ \eta-b\right] \,\bar{U}^{\alpha}\,\left(\bar{U}_{\mu}\bar{U}^{\mu}\right).
\end{equation}
After some algebra, can be obtained that \cite{p2}
\begin{eqnarray}
\left<B\right|\hat{\Omega}_{\alpha}\left(x^{\mu}\right)\left|B\right> &=& \bar{U}_{\alpha}, \quad \left<B\right|\hat{\Omega}^{\alpha}\left(x^{\mu}\right)\left|B\right> = \bar{U}^{\alpha}, \label{asa}\\
\left<B\right|\hat{\Omega}_{\alpha}\left(x^{\mu}\right)\,\hat{\Omega}^{\alpha}\left(x^{\mu}\right)\left|B\right>&=& \left[\bar{\theta}(z)-1\right], \label{asa1}
\end{eqnarray}
where $\bar{\theta}(z)$ is \cite{p2}
\begin{equation}\label{teta}
\bar{\theta}(z)= \frac{4z^3-3z^2-10z+8+\sqrt{9z^2-8z}(2-z)}{z(z-1)\left[\sqrt{9z^2-8z}-z\right]}.
\end{equation}
The quantum expectation value of the quantum varied connection (\ref{cone}), must be the classical varied connection of the manifold:
\begin{equation}
\left<B\right|\hat{\delta\Gamma}^{\alpha}_{\mu\nu}\left|B\right> = {\delta\Gamma}^{\alpha}_{\mu\nu}= b \,\bar{U}^{\alpha}\, g_{\mu\nu}.
\end{equation}

Since we have obtained the covariant metric tensor that includes the effects of the fluctuations: $\bar{g}_{\alpha\beta}$, it will now be possible to calculate the contravariant components $\bar{g}^{\alpha\beta}$ and construct a standard relativistic formalism without boundary terms in the varied action (\ref{delta0}), such that these terms are included in the line element. This is what we will do in the next section, and then illustrate the formalism by developing an inflationary model starting from a massive scalar field (inflaton). We will study its dynamics on the new Riemannian manifold and will try to quantize the field taking into account the quantum gravity effects due to the inflaton field, without using the semiclassical approximation.

\section{New GR with quantum gravitational effects included in the background geometry}

Once incorporated, the quantum gravitational spacetime fluctuations in the metric tensor, we can describe our new Einstein equations on the background metric $\bar{g}_{\alpha\beta}$, after variating the EH action $\bar{{\cal I}} = \int\,d^4x\,\sqrt{-\bar{g}}\,\left\{\frac{\bar{R}}{2\kappa}+\left<V\right|\hat{\cal L}_m\left|V\right>\right\}$:
\begin{equation}\label{delta1}
\delta \bar{{\cal I}} = \int d^4 x \sqrt{-\bar{g}} \left[{\delta \bar{g}}^{\alpha\beta}\,\bar{G}_{\alpha\beta}+  \left<V\right|{\delta g}^{\alpha\beta}  \kappa\, \hat{T}_{\alpha\beta}\left|V\right>+ \,\bar{g}^{\alpha\beta} {\delta \bar{R}}_{\alpha\beta} \right]=0,
\end{equation}
where the quantum state $\left|V\right>$ is in the Heisenberg representation in which quantum operators evolve in spacetime, while the quantum states are squeezed in space and time. Notice that $\left|B\right>$ acts on the background described by $g_{\alpha\beta}$, but the states $\left|V\right>$ act on the new background with quantum fluctuation included, which is described by $\bar{g}_{\alpha\beta}$, being related to a zero net flux:
\begin{equation}
\bar{g}^{\alpha\beta}\,{\delta \bar{R}}_{\alpha\beta} =0,
\end{equation}
where the varied Ricci tensor, now is defined by using the extended Palatini expression \cite{pal}, with respect to the Levi-Civita connections $\bar{\Gamma}^{\mu}_{\alpha\beta}$, which we shall denote as $\left(...\right)_{;}$:
\begin{equation}\label{ricci}
{\delta \bar{R}}_{\alpha\beta} = b\,\left[\left({ \delta\bar{\Gamma}}^{\mu}_{\alpha\mu} \right)_{; \beta} - \left({ \delta\bar{\Gamma}}^{\mu}_{\alpha\beta}\right)_{; \mu}\right].
\end{equation}
For ${ \delta\bar{\Gamma}}^{\alpha}_{\mu\nu}=\bar{b}\,\bar{\Omega}^{\alpha}\,\bar{g}_{\mu\nu}$, the equations of motion for $\bar{\Omega}^{\alpha}$, will be
\begin{equation}
\left(\bar{\Omega}^{\alpha}\right)_{; \alpha}= \bar{g}^{\alpha\beta}\,\left(\bar{\Omega}_{\alpha}\right)_{; \beta}=0.
\end{equation}
The resulting equation of motion for $\bar{\Omega}$ is
\begin{equation}\label{flu}
\bar{\Box} \bar{\Omega}=0.
\end{equation}
The components of the relativistic velocity in the representation without flow: $u^{\alpha}=\frac{dx^{\alpha}}{\hskip -.15cm ds}$, comply with the normalization (\ref{newv}), and are solutions of the geodesic equation without sources:
\begin{equation}\label{ge}
\frac{du^{\alpha}}{ds} + \bar{\Gamma}^{\alpha}_{\mu\nu}\, u^{\mu}u^{\nu}=0,
\end{equation}
such that the new Levi-Civita connections $\bar{\Gamma}^{\alpha}_{\mu\nu}$ are given by
\begin{equation}\label{levi}
\bar{\Gamma}^{\alpha}_{\mu\nu} = \frac{1}{2} \bar{g}^{\alpha\epsilon} \left[\bar{g}_{\mu\epsilon,\nu}+ \bar{g}_{\nu\epsilon,\mu} - \bar{g}_{\mu\nu,\epsilon}\right].
\end{equation}
The equations (\ref{flu}) and (\ref{ge}) mean that the relativistic observer will move in a co-moving reference frame with the tidal wave of dark energy associated with the flow of spacetime fluctuations. In this framework we can consider the Einstein equations without sources, by considering the tensor metric components (\ref{gbar}), and the Levi-Civita connections (\ref{levi}):
\begin{equation}\label{eins}
\bar{G}_{\mu\nu} = - \kappa\,\left<B\right|\bar{T}_{\mu\nu}\left|B\right>,
\end{equation}
where $\bar{G}_{\mu\nu}=\bar{R}_{\mu\nu}-(1/2)\,\bar{R}\,\bar{g}_{\mu\nu} $ is the Einstein tensor. The curvature tensor is defined by
\begin{equation}
\bar{R}^{\alpha}_{\,\,\,\beta\gamma\delta} = \bar{\Gamma}^{\alpha}_{\,\,\,\beta\delta,\gamma} -  \bar{\Gamma}^{\alpha}_{\,\,\,\beta\gamma,\delta}
+ \bar{\Gamma}^{\epsilon}_{\,\,\,\beta\delta} \bar{\Gamma}^{\alpha}_{\,\,\,\epsilon\gamma} - \bar{\Gamma}^{\epsilon}_{\,\,\,\beta\gamma}
\bar{\Gamma}^{\alpha}_{\,\,\,\epsilon\delta},
\end{equation}
and the Ricci tensor is $\bar{R}_{\beta\gamma}=\bar{R}^{\alpha}_{\,\,\,\beta\gamma\alpha}$. In other words, the new Einstein formalism with
quantum fluctuations included can be developed on a new Riemann manifold, where the metric tensor components (\ref{gbar}) constitute the
fundamental tensor with covariant null covariant derivative [which is calculated with respect to the Levi-Civita connections (\ref{levi})]: $\bar{g}_{\mu\nu ;\alpha}=0$. Furthermore, the stress tensor $\bar{T}_{\mu\nu}$ is defined as
\begin{equation}
\bar{T}_{\mu\nu}= 2 \frac{\delta \hat{\cal L}_m}{\delta \bar{g}^{\mu\nu}} -  \bar{g}_{\mu\nu}\,   \hat{\cal L}_m.
\end{equation}
We shall consider that the expectation value of the stress tensor, calculated on the background metric (\ref{gbar}), is described by a perfect fluid, such that
\begin{equation}
\left<B\right|\bar{T}^{\mu\nu}\left|B\right> = \left(P+\rho\right) u^{\mu}u^{\nu} - \bar{g}^{\mu\nu} \,P,
\end{equation}
where for $\bar{T}=\bar{g}_{\mu\nu}\,\bar{T}^{\mu\nu}$, we obtain that $\left<B\right|\bar{T}\left|B\right>= \rho-3P $. The equation of state for the system is given by $\varpi=P/\rho$:
\begin{equation}
\varpi = \frac{P}{3P+\left<V\right|\bar{T}\left|V\right>}.
\end{equation}
Furthermore, the field equations take the form
\begin{equation}
\left(\bar{G}^{\alpha\beta}\right)_{;\alpha} =  \,\left<B\right|\left[\bar{T}^{\alpha\beta}\right]_{;\alpha}\left|B\right>=0,
\end{equation}
in agreement with that one expects in Standard GR in absence of external sources. We must remember that the covariant derivative $\left[...\right]_{;\alpha}$, is
defined with respect to the Levi-Civita connections $\bar{\Gamma}^{\alpha}_{\mu\nu}$, of Eq. (\ref{levi}).

\section{Cosmic inflation with nonperturbative quantum gravity included}

To illustrate the above formalism, we shall consider a particular case of cosmic inflation \cite{in0,in1,in11,in2,in3,in33,in333,hm,in4,b1,bcms,vega1,maube1,maube2,vega2,vega3,vega4,mrv}.
In the inflationary epoch, the energy density of the universe was overwhelmingly dominated by the inflaton field, which is a massive scalar field, which was characterized by a negligible kinetic energy density. The resultant vacuum energy density was directly responsible for the exponential expansion of the cosmic scale factor. During this phase, an initially compact and causally connected spacetime region, on the order of the Hubble radius, underwent immense super-horizon growth, thereby readily encompassing the entirety of the presently observed Universe's comoving volume. This mechanism can explain the observed spatial homogeneity and isotropy of the Universe. Moreover, it is now firmly established that cosmic structure formation fundamentally stems from an almost scale-invariant, super-horizon curvature perturbation. Inflationary cosmology thus provides a robust framework for understanding the early quasi-exponential expansion of the universe and the quantum to classical transition of the initially quantum fluctuations of the inflaton field \cite{np} at macroscopic scales. In our model, we shall consider that the parameter $\eta$ is given by $\eta=-\frac{H_0}{(z-1)\left[\theta(z)-1\right]}$.

\subsection{Generalities}

To describe the inflationary model we shall consider an universe which is considered isotropic, but not homogeneous [the spatial curvature is considered negative: $K=-H^2_0$]. We shall consider a Lagrangian density for a massive quantum scalar (inflaton) field $\hat{\varphi}$ which is minimally coupled to gravity, but that evolves on a Riemann manifold which includes the quantum gravity fluctuations
\begin{equation}\label{lagrange}
\hat{\cal L}_m\left(\hat{\varphi}, \hat{\varphi}_{,\mu}\right) = \frac{1}{2}\, \bar{g}^{\alpha\beta} \hat{\varphi}_{,\alpha}\hat{\varphi}_{,\beta} - \frac{{\cal M}^2}{2}\,\hat{\varphi}^2,
\end{equation}
where $\hat{\varphi}_{,\alpha}\equiv \partial_{\alpha}\hat{\varphi}$, and $\bar{g}^{\alpha\beta}$ is represented by the adjoint matrix of $\bar{g}_{\alpha\beta} $, given by (\ref{covg}):
\begin{equation}
\bar{g}^{\alpha\beta} = \frac{\left[\bar{g}_{\alpha\beta}\right]^{\dagger}}{\left|\bar{g}\right|},
\end{equation}
where $\bar{g}\equiv \det{\left[\bar{g}_{\alpha\beta}\right]}$, is the determinant of the covariant metric tensor $\bar{g}_{\alpha\beta}$, so that $\bar{g}_{\alpha\beta}\,\bar{g}^{\alpha\beta}=4$ and
\begin{equation}\label{newv}
\bar{g}_{\alpha\beta}\,{u}^{\alpha}\,{u}^{\beta}=1.
\end{equation}
In our model of inflation, we shall consider that the relativistic observer is in a co-moving frame on the background spacetime, so that we can take $\bar{U}^0\neq 0$, and $\bar{U}^i=0$, in the equations. (\ref{geo}). The line element with quantum fluctuations included will be described by
\begin{equation}
ds^2 = \bar{g}_{00} \,d\tau^2 +\bar{g}_{11} \,dr^2 + \left[\bar{g}_{22}\,d\vartheta^2 + \bar{g}_{33} \,d\phi^2 \right],
\end{equation}
where, for a negative spatial curvature $K=-H^2_0$, the relevant components of the metric tensor with quantum gravity effects included, are
\begin{eqnarray}\label{gbar}
\bar{g}_{00} &= & \left[1+\frac{2\left[\theta(z)-1\right]}{z} \right] -2\,\bar{U}^2_{0}, \qquad
\bar{g}_{11} = \frac{e^{2H_0\,\tau}}{1+H^2_0\,r^2}\,\left[1+\frac{2\left[\theta(z)-1\right]}{z} \right] , \label{gbar} \\
\bar{g}_{22} &= & e^{2H_0\,\tau}\,r^2\,\left[1+\frac{2\left[\theta(z)-1\right]}{z} \right] , \qquad
\bar{g}_{33} = e^{2H_0\,\tau}\,r^2\,\sin^2(\vartheta)\,\left[1+\frac{2\left[\theta(z)-1\right]}{z} \right],
\end{eqnarray}
where $\theta(z)$ is given by (\ref{oro}), and $H_0$ is the cosmological constant. Furthermore, $\bar{U}_{0}={g}_{00} \,\bar{U}^0$ and $\bar{U}^0$ is the solution of Eq. (\ref{geo}), for a line element without fluctuations included and $g_{\mu\nu}\,\bar{U}^{\mu}\bar{U}^{\nu}=\theta(z)-1$:
\begin{equation}\label{U0}
\bar{U}_{0}(\tau) = \sqrt{\frac{\theta(z)-1}{2}}\, e^{\eta(z-1)\left[\theta(z)-1\right]\tau},
\end{equation}
where the spacetime without fluctuations is described by $ds^2= g_{\alpha\beta}\,dx^{\alpha}dx^{\beta}$. Will be reasonable to consider a frame with $u^0(\tau,r)\neq 0$, $u^1(\tau,r)\neq 0$ and $u^2=u^3=0$. The normalization condition for the relativistic velocities takes the form: $\bar{g}^{00} u_0\,u_0+\bar{g}^{11} u_1\,u_1=1$. We shall assume that $\eta=-\frac{H_0}{(z-1)\left[\theta(z)-1\right]}$, for $z\neq 1$ and $z\neq 0$. The relevant Levi-Civita connections
are:
\begin{eqnarray}
&& \bar{\Gamma}^{0}_{\hskip.1cm 00} = -\frac{z\left[\theta(z)-1\right]\,H_0\,e^{-2H_0\tau}}{\left[\theta(z)-1\right]\left[ze^{-2H_0\tau}-2 \right]-z}, \qquad \bar{\Gamma}^{1}_{\hskip.1cm 01} = H_0,   \\
&& \bar{\Gamma}^{0}_{\hskip.1cm 11} =- \frac{H^2_0 e^{2H_0\tau} \left\{z+2\left[\theta(z)-1\right]\right\}}{\left[1+H^2_0r^2\right]\left\{\left[\theta(z)-1\right]\left[\right]\left[ze^{-2H_0\tau}-2 \right]-z\right\}}, \qquad \bar{\Gamma}^{1}_{\hskip.1cm 11} =-\frac{H_0 \,r}{\left[1+H^2_0r^2\right]}.
\end{eqnarray}
The nonzero contravariant components of the relativistic velocity for the geodesic equations ($\ref{ge}$), are:
\begin{eqnarray}
u^0(\tau,r) &=& \pm \frac{e^{H_0\tau} \sqrt{\left\{e^{2H_0\tau}\left\{z+2\left[\theta(z)-1\right]\right\}-z\left[\theta(z)-1\right]\right\}\left\{2C^2e^{-2H_0\tau}\theta(z)+
C^2e^{-2H_0\tau} +z\right\}}}{\left\{e^{2H_0\tau} \left\{z+2\left[\theta(z)-1\right]\right\}-z\left[\theta(z)-1\right]\right\}}, \label{v0} \\
u^1(\tau,r) &=& C\,\left[1+H^2_0 r^2\right]^{1/2} \,e^{-2H_0\tau}. \label{v1}
\end{eqnarray}
Notice that when the constant $C=0$, it is obtained that $\left.u^0(\tau,r)\right|_{C=0} \neq 0 $ and $\left.u^1(\tau,r)\right|_{C=0} = 0 $. The scalar curvature $\bar{R} = \bar{g}^{\alpha\beta} \bar{R}_{\alpha\beta}$, is given by
\begin{eqnarray}
\bar{R} &=& -\frac{6 z^{3} \left[\theta \left(z \right)-1\right]^{2} {H_0}^{2} {e}^{-2 {H_0} \tau}}{\left\{z \left[\theta\left(z \right)-1\right]-\left[z +2 \left[\theta \left(z \right)-1\right]\right]{e}^{2 H_0 \tau}\right\}^{2} \left[z +2 \left[\theta \left(z \right)-1\right]\right]} \nonumber \\
&-&\frac{6 z \left\{3 \left[z +2 \left[\theta \left(z \right)-1\right]\right] \left[\left(z +\frac{2}{3}\right) \theta \left(z \right)-\frac{2 }{3}(z+1)\right] {e}^{2 H_0 \tau}-2 \left[z +2 \left[\theta \left(z \right)-1\right]\right]^{2} {e}^{4 H_0 \tau}\right\}\, H_0^{2}}{\left\{z \left[\theta\left(z \right)-1\right]-\left[z +2 \left[\theta \left(z \right)-1\right]\right]{e}^{2 H_0 \tau}\right\}^{2} \left[z +2 \left[\theta \left(z \right)-1\right]\right]} \nonumber \\
&-& \frac{6 z \left\{2\left[z +2 \left[\theta \left(z \right)-1\right]\right] z \left[\theta \left(z \right)-1\right]\right\} H_0^{2}}{\left\{z \left[\theta\left(z \right)-1\right]-\left[z +2 \left[\theta \left(z \right)-1\right]\right]{e}^{2 H_0 \tau}\right\}^{2} \left[z +2 \left[\theta \left(z \right)-1\right]\right]}.
\end{eqnarray}
Notice that when $\tau \rightarrow \infty$, the scalar curvature tends to
\begin{equation}
\left.\bar{R}\right|_{\tau\rightarrow\infty} \rightarrow \frac{12 \,z}{\left[z +2 \left[\theta \left(z \right)-1\right]\right]} \,H^2_0.
\end{equation}
For $\eta=-\frac{H_0}{(z-1)\left[\theta(z)-1\right]}$ in Eqs. (\ref{gbar}) and (\ref{U0}), the equation of state ${\varpi} = -\frac{\bar{G}^1_{\hskip .1cm 1}}{\bar{G}^0_{\hskip .1cm 0}}\equiv \frac{P}{\rho}$, take the following particular form
\begin{eqnarray}
\varpi &=& \frac{e^{2H_0\tau}\left[12\left[\theta^2(z)+\theta(z)\,z-z+1\right]-3\left[z^2+8\theta(z)\right]\right]-4\left[\theta(z)+1\right]^2+4z^2+16z\theta(z)-6z-5z\theta(z)
\left[2\theta(z)+z\right]}{3\,e^{2H_0\tau}\left[2\left[\theta(z)-1\right]+z\right]\left[\left[2\left[\theta(z)-1\right]+z\right]-2 e^{-2H_0\tau}\left[\left[\theta(z)-1\right]+z\right]+2\left[\theta(z)-1\right]e^{-4H_0\tau}\right]} \nonumber \\
&-& \frac{e^{-2H_0\tau}z\left[z-4-2\theta(z)\left[2\theta(z)+z-4\right]\right] +
e^{-4H_0\tau}\left[z^2\left[\theta^2(z)+2\theta(z)-1
\right]\right]}{3\,e^{2H_0\tau}\left[2\left[\theta(z)-1\right]+z\right]\left[\left[2\left[\theta(z)-1\right]+z\right]-2 e^{-2H_0\tau}\left[\left[\theta(z)-1\right]+z\right]+2\left[\theta(z)-1\right]e^{-4H_0\tau}\right]}.
\end{eqnarray}

\begin{figure}
    \centering
    \includegraphics[scale=0.6]{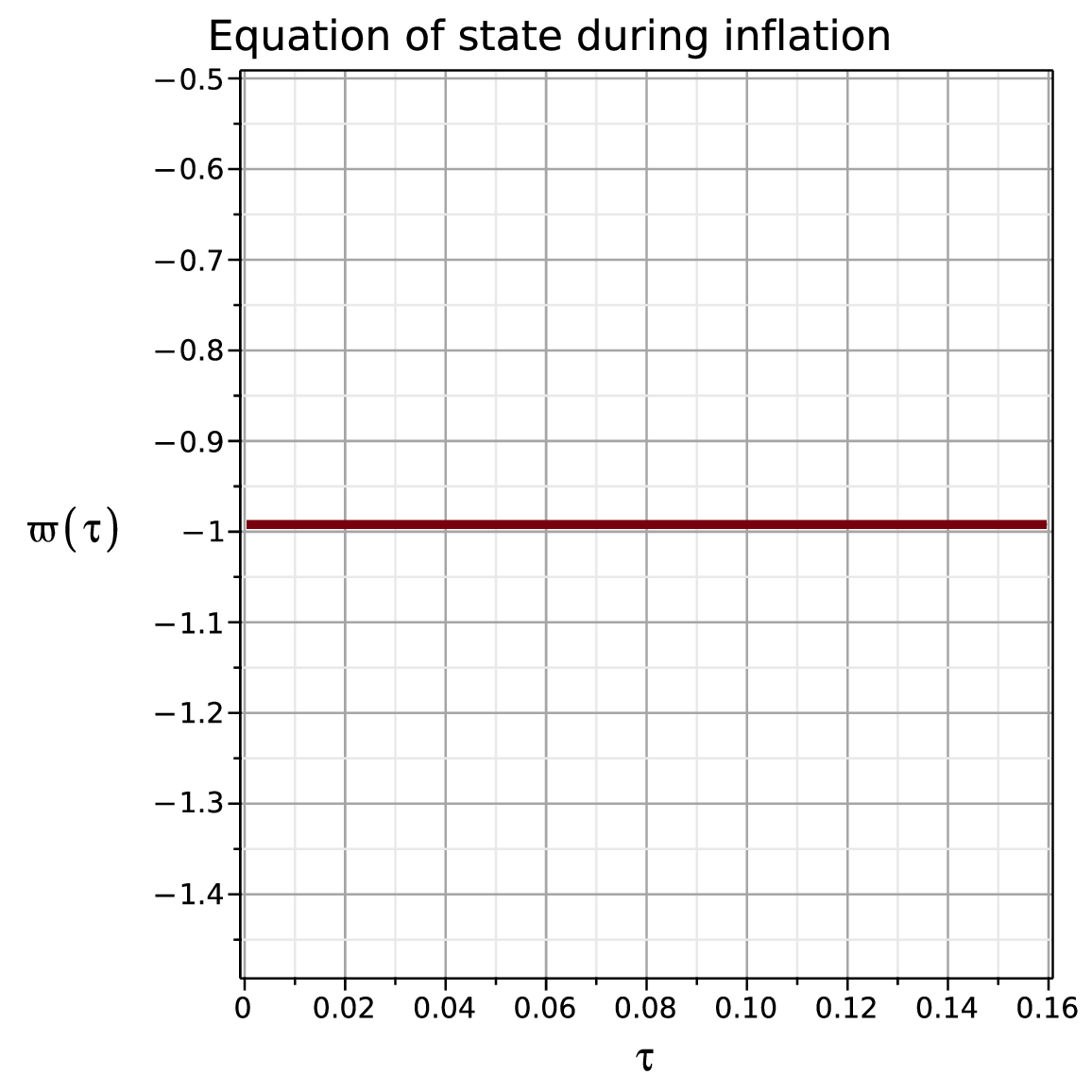}
    \caption{Evolution of $\varpi(\tau)$ for $z=0.998$ and $0<\tau < 1.6\times 10^{-1}\,{\rm meters}$, that remains constant during the early inflationary epoch: $\varpi \simeq -0.991969$.}
    \label{f1gi148}
\end{figure}

In the Fig. (\ref{f1gi148}) we show the evolution during inflation of $\varpi$ for $z=0.998$, and $H_0= 7.452587438\times 10^{-27}\,({\rm met.})^{-1}$ [which corresponds to the value $H_0=67.74\,{\rm km/(seg. Mpc)} $], as a function of $\tau$. For this $z$-value the large-scale quantum correlations of spacetime are relevant and $\left<V\right|\hat{\Omega}^{\alpha}\hat{\Omega}_{\alpha}\left|V\right> \neq \left<V\right|\hat{\Omega}^{\alpha}\left|V\right>\left<V\right|\hat{\Omega}_{\alpha}\left|V\right>={u}^{\alpha}{u}_{\alpha}=1$. Notice that the equation of state remains always constant with a value $\varpi = -0.991969 $.

\subsection{Dynamics of geometric and physical scalar fields: $\hat{\Omega}$ and $\hat{\varphi}$}

To describe the dynamics of the system we must consider the dynamics of scalar fields which describe the geometric quantum dynamics of spacetime and the physical system, on the background spacetime described by the Riemann manifold in which the metric $\bar{g}_{\alpha\beta}$ is the fundamental tensor. These fields are $\hat{\Omega}$ and $\hat{\varphi}$.
In this framework, $\hat{\varphi}$ is a massive scalar field that is responsible for the curvature of spacetime in the early universe. We consider the lagrangian density: $\hat{\cal L}_m\left(\hat{\varphi}, \hat{\varphi}_{,\mu}\right)$ given by (\ref{lagrange}). The dynamics of $\hat{\varphi}$ will be given by
\begin{equation}\label{el}
\partial_{\mu} \frac{\delta \hat{L}}{\delta \hat{\varphi}_{,\mu}}- \frac{\delta \hat{L}}{\delta \hat{\varphi}}=0,
\end{equation}
where $\hat{L}= \sqrt{-\bar{g}}\, \hat{\cal L}_m\left(\hat{\varphi}, \hat{\varphi}_{,\mu}\right)$. The equation (\ref{el}) takes the form
\begin{equation}\label{varphi}
\left(\frac{1}{\sqrt{-\bar{g}}}\, \frac{\partial}{\partial x^{\epsilon}}\sqrt{-\bar{g}}\right)\,\bar{g}^{\alpha\epsilon} \hat{\varphi}_{,\alpha} + \bar{g}^{\alpha\epsilon}_{,\epsilon}\, \hat{\varphi}_{,\alpha} + \bar{g}^{\alpha\epsilon} \hat{\varphi}_{,\alpha\epsilon} + \frac{\delta \hat{V}(\hat{\varphi})}{\delta\,\hat{\varphi}} =0.
\end{equation}
On the other hand, the resulting equation of motion for $\hat{\Omega}$ is given by the flux equation (\ref{flu}), which takes the form
\begin{equation}\label{omega}
\left(\frac{1}{\sqrt{-\bar{g}}}\, \frac{\partial}{\partial x^{\epsilon}}\sqrt{-\bar{g}}\right)\,\bar{g}^{\alpha\epsilon} \hat{\Omega}_{,\alpha} + \bar{g}^{\alpha\epsilon}_{,\epsilon}\, \hat{\Omega}_{,\alpha} + \bar{g}^{\alpha\epsilon} \hat{\Omega}_{,\alpha\epsilon}=0,
\end{equation}
where we must remember that $\hat{\Omega}_{,\mu}\equiv \hat{\Omega}_{\mu}$.

\subsection{Quantization of the inflaton field with geometric fluctuations of spacetime}

To realize the quantization during inflation taking into account the geometric quantum fluctuations of spacetime, we can consider the Fourier expansions for both, the scalar inflaton field $\hat{\varphi}$, and the scalar geometric field $\hat{\Omega}$
\begin{eqnarray}
\hat{\varphi}\left(\tau,\vec{r}\right) &=& \int^{\infty}_{0} dk\,k^2\,\sum_{lm} \left[\hat{C}_{klm}
\varphi_{klm}(\vec r,\tau) +\hat{C}^{\dagger}_{klm} \varphi^*_{klm}(\vec
r,\tau)\right] , \label{f1} \\
\hat{\Omega}\left(\tau,\vec{r}\right) & = & \int^{\infty}_{0} dk\,k^2\,\sum_{lm} \left[\hat{A}_{klm}
\Omega_{klm}(\vec r,\tau) +\hat{A}^{\dagger}_{klm} \Omega^*_{klm}(\vec
r,\tau)\right], \label{f2}
\end{eqnarray}
where, $\hat{C}^{\dagger}_{klm}$ and $\hat{C}_{klm}$ are respectively the creation and annihilation operators of the massive inflaton field and
$\hat{A}^{\dagger}_{klm}$ and $\hat{A}_{klm}$ are respectively the creation and annihilation operators for the geometric field $\hat{\Omega}$.
Furthermore, $\sum_{lm}\equiv \sum_{l=0}^{\infty}\, \sum_{m=-l}^{l}$, $k$, $l$ and $m$ are respectively the wave-numbers related to the coordinates $r$, $\vartheta$ and $\phi$, and the modes for both scalar fields are
\begin{equation}\label{sep}
\varphi_{klm}(\vec r,\tau)= R_{kl}\left(r\right)\,
Y_{lm}(\vartheta,\phi) \,\xi_k(\tau), \qquad \Omega_{klm}(\vec r,\tau)= R_{kl}\left(r\right)\,
Y_{lm}(\vartheta,\phi) \,\chi_k(\tau),
\end{equation}
Here, $Y_{lm}(\vartheta,\phi)$ are the spherical harmonics. Furthermore, the annihilation and
creation operators obey the algebra
\begin{eqnarray}
&&\left<V\right| \left[\hat{A}_{klm}, \hat{A}^{\dagger}_{k'l'm'}\right]\left|V\right> = {1\over
(2\,k)^{3}}\,\frac{(2l+1)(l-m)!}{4\pi (l+m)!}\,
\delta({k}-{k'})\, \delta_{ll'}\, \delta_{m m'}, \\
&&\left<V\right|\left[\hat{A}_{klm}, \hat{A}_{k'l'm'}\right]\left|V\right>=\left<V\right|\left[\hat{A}^{\dagger}_{klm},
\hat{A}^{\dagger}_{k'l'm'}\right]\left|V\right>=0,  \label{cc1} \\
&& \left<V\right|\left[\hat{C}_{klm}, \hat{C}^{\dagger}_{k'l'm'}\right] \left|V\right>={1\over
(2\,k)^{3}}\,\frac{(2l+1)(l-m)!}{4\pi (l+m)!}\,
\delta({k}-{k'})\, \delta_{ll'}\, \delta_{m m'}, \\
&&\left<V\right|\left[\hat{C}_{klm}, \hat{C}_{k'l'm'}\right]\left|V\right>=\left<V\right|\left[\hat{C}^{\dagger}_{klm},
\hat{C}^{\dagger}_{k'l'm'}\right]\left|V\right>=0, \label{cc2}
\end{eqnarray}
with $\hat{A}_{klm}={1\over
(2\,k)^{3/2}}\,\sqrt{\frac{(2l+1)(l-m)!}{4\pi (l+m)!}}$ and $\hat{C}_{klm}={1\over
(2\,k)^{3/2}}\,\sqrt{\frac{(2l+1)(l-m)!}{4\pi (l+m)!}}$. To make possible a quantum relativistic description of quantum fields we must keep in mind that the commutator $\left[\hat{\varphi}(x),\hat{\Pi}^{\mu}(x') \right]$ should give us as a result a geometrical quantum field \cite{p2}:
\begin{eqnarray}
&& \left[\hat{\varphi}(\tau,\vec{r}),\hat{\Pi}^{\mu}(\tau,\vec{r}') \right] = -i\,\hat{\Omega}^{\mu}(\tau,\vec{r}')\,\delta^{(3)}\left(\vec{r}-\vec{r}'\right), \label{cua1}\\
&& \left[\hat{\varphi}(\tau,\vec{r}),\hat{\Pi}_{\mu}(\tau,\vec{r}'') \right] = i\,\hat{\Omega}_{\mu}(\tau,\vec{r}'')\,\delta^{(3)}\left(\vec{r}-\vec{r}''\right), \label{cua2}\\
&& \left[\hat{\varphi}(\tau,\vec{r}),\hat{\Pi}_{\mu}(\tau,\vec{r}') \right] \left[\hat{\varphi}(\tau,\vec{r}),\hat{\Pi}^{\mu}(\tau,\vec{r}'') \right] =
\hat{\Omega}_{\mu}\left(\tau,\vec{r}'\right)\hat{\Omega}^{\mu}(\tau,\vec{r}'')\,\delta^{(3)}(\vec{r}-\vec{r}')\,
\delta^{(3)}\left(\vec{r}-\vec{r}''\right), \label{cua3}
\end{eqnarray}
where the momentum is given by $\hat{\Pi}^{\mu}= \frac{\delta \hat{{\cal L}}_m}{\delta \hat{\varphi}_{,\mu}}$. From Eqs. (\ref{asa}) and (\ref{asa1}), we obtain
\begin{eqnarray}
&&\left<V\right|\left[\hat{\varphi}(\tau,\vec{r}),\hat{\Pi}^{\mu}(\tau,\vec{r}') \right]\left|V\right> =- i\,{u}^{\mu}(\tau,\vec{r}')\,\delta^{(3)}\left(\vec{r}-\vec{r}'\right), \label{c1} \\
&&\left<V\right|\left[\hat{\varphi}(\tau,\vec{r}),\hat{\Pi}_{\mu}(\tau,\vec{r}'') \right]\left|V\right> = i\,{u}_{\mu}(x'')\,\delta^{(3)}\left(\vec{r}-\vec{r}''\right), \label{c2} \\
&&\left<V\right|\left[\hat{\varphi}(\tau,\vec{r}),\hat{\Pi}_{\mu}(\tau,\vec{r}') \right] \left[\hat{\varphi}(\tau,\vec{r}),\hat{\Pi}^{\mu}(\tau,\vec{r}'') \right]\left|V\right> = \nonumber \\
&& \left<V\right|\hat{\Omega}_{\mu}\left(\tau,\vec{r}'\right)\hat{\Omega}^{\mu}(\tau,\vec{r}'')\left|V\right>\,\delta^{(3)}(\vec{r}-\vec{r}')\,
\,\delta^{(3)}(\vec{r}-\vec{r}')\,\delta^{(3)}\left(\vec{r}-\vec{r}''\right), \label{c3}
\end{eqnarray}
which are the equal time expectation values on the background Riemannian manifold described by the metric tensor components
$\bar{g}_{\alpha\beta}$. Notice that we have used the fact that $\bar{g}_{\alpha\beta} \,{u}^{\alpha}{u}^{\beta}=1$. Notice that (\ref{c1}) and (\ref{c2}) are dependent on the relativistic frame, but not (\ref{c3}), which is an invariant for a given $z$-value. Notice that we have required that validity of the following non-linear quantum algebra on the new background:
\begin{equation}\label{alg}
\left[\frac{\hat{\partial}}{\partial{x}^{\alpha}},\hat{\Omega}_{k'}^\dagger\right]=\,{u}_{\alpha}\,
\left[\hat{\Omega}_{k},\hat{\Omega}_{k'}^\dagger\right],
\end{equation}
where $\hat{\Omega}_{klm}(\tau,\vec{r})=\hat{A}_{klm}\,{\Omega}_{klm}(\tau,\vec{r})$ and $\hat{\Omega}^{\dagger}_{klm}(\tau,\vec{r})=\hat{A}^{\dagger}_{klm}\,{\Omega}^*_{klm}(\tau,\vec{r})$, and
$\left<V\right|\hat{\Omega}_{\mu}\left|V\right>={u}_{\mu}$.

\subsection{Exact solutions for the modes with $\eta=-\frac{H_0}{(z-1)\left[\theta(z)-1\right]}$}

To find the solutions of the dynamic equations (\ref{varphi}) and (\ref{omega}), we shall consider the separation of variables in Eq. (\ref{sep}), for both scalar fields, $\hat{\varphi}$ and $\hat{\Omega}$. The spatial solutions are the same for both fields. However, the dynamics of $\xi_k(\tau)$ and $\chi_k(\tau)$ are different. The differential equation that describes the dynamics of $\xi_k(\tau)$, for $\eta=-\frac{H_0}{(z-1)\left[\theta(z)-1\right]}$, is given by
\begin{eqnarray}
\frac{d^2\xi_k}{d\tau^2} &+& \frac{\left[\left[\theta(z)-1\right]H_0 e^{-2H_0\tau}z+3H_0\left[\left[\theta(z)-1\right]e^{-2H_0\tau}z-2\left[\theta(z)-1\right]-z\right]\right]}{\left[\left[\theta(z)-1\right]ze^{-2H_0\tau}-2\left[\theta(z)-1\right]-z
\right]} \frac{d\xi_k}{d\tau} \nonumber \\
&-& \left[\frac{\left[\left[zk^2e^{-2H_0\tau}+{\cal M}^2\left[z+2\left[\theta(z)-1\right]\right]\right]\left[\left[\theta(z)-1\right]ze^{-2H_0\tau}-2\left[\theta(z)-1\right]-z\right]^2\right]}{\left[z+2\left[\theta(z)-1\right]\right]z\left[
\left[\theta(z)-1\right]ze^{-2H_0\tau} - 2 \left[\theta(z)-1\right]-z\right]}\right]\,\xi_k(\tau)=0,
\end{eqnarray}
which has the general solution
\begin{equation}\label{xi}
\xi_k(\tau) = C_1\, {\cal H_C}\left[\alpha_1,\beta_1,-\frac{3}{2},\gamma_1,\delta_1,f(\tau)\right]\,e^{h_1(k,\tau)} + i\,C_2\, {\cal H_C}\left[\alpha_1,-\beta_1,-\frac{3}{2},\gamma_1,\delta_1,f(\tau)\right]\,e^{-h_1(k,\tau)},
\end{equation}
where $C_1, C_2$ are constants to be determined and the parameters of the confluent Heun functions: ${\cal H_C}$, and
\begin{eqnarray}
\alpha_1 & = & \frac{k}{H_0\sqrt{z}}\sqrt{\frac{2\,\left[\theta(z)-1\right]+z }{\left[\theta(z)-1\right]}}, \qquad
\beta_1 = -\frac{\sqrt{9H^2_0z-4{\cal M}^2\left[2\left[\theta(z)-1\right]+z\right]}}{2\sqrt{z}H_0}, \\
\gamma_1 &=& \frac{\left[k^2-{\cal M}^2\left[\theta(z)-1\right]\right]\left[2\left[\theta(z)-1\right]+z\right]}{4z\,H^2_0}, \qquad f(\tau) = \frac{z\left[\theta(z)-1\right]\,e^{-2H_0\tau}}{2\left[\theta(z)-1\right]+z}, \\
\delta_1 &=& \frac{4{\cal M}^2 \theta^2(z)+ \left[2{\cal M}^2 (z-4)+3H^2_0 z-4k^2\right]\theta(z)+2(2-z) {\cal M}^2-3H^2_0 z-2k^2(z-2)}{8 \left[\theta(z)-1\right]z H^2_0}, \\
h_1(k,\tau) & = & \frac{\left[z^{3/2}+2\sqrt{z} \left[\theta(z)-1\right]\right]H_0\tau \sqrt{9H^2_0z-4{\cal M}^2\left[2\left[\theta(z)-1\right]+z\right]}  -3zH^2_0\tau \left[2\left[\theta(z)-1\right]+z\right]}{2z H_0 \left[z+2\left[\theta(z)-1\right]\right]} \nonumber \\
&+& \frac{k \,z^{3/2} e^{-2H_0\tau} \left[\theta(z)-1\right] }{2z H_0 \left[z+2\left[\theta(z)-1\right]\right]}\sqrt{\frac{2\left[\theta(z)-1\right]+z}{\left[\theta(z)-1\right]}}.
\end{eqnarray}
On the other hand, the equation of motion for $\chi_k(\tau)$ is given by
\begin{eqnarray}
\frac{d^2\chi_k}{d\tau^2} &+& \frac{\left[\left[\theta(z)-1\right]H_0 e^{-2H_0\tau}z+3H_0\left[\left[\theta(z)-1\right]e^{-2H_0\tau}z-2\left[\theta(z)
-1\right]-z\right]\right]}{\left[\left[\theta(z)-1\right]ze^{-2H_0\tau}-2\left[\theta(z)-1\right]-z
\right]} \frac{d\chi_k}{d\tau} \nonumber \\
&-& \left[\frac{\left[\left[zk^2e^{-2H_0\tau}\right]\left[\left[\theta(z)-1\right]ze^{-2H_0\tau}-2\left[\theta(z)-1\right]-z\right]^2\right]}{\left[z+2\left[\theta(z)-1\right]\right]z\left[
\left[\theta(z)-1\right]ze^{-2H_0\tau} - 2 \left[\theta(z)-1\right]-z\right]}\right]\,\chi_k(\tau)=0,
\end{eqnarray}
which has the general solution
\begin{equation}\label{chi}
\chi_k(\tau) = C_3\, {\cal H_C}\left[\alpha_2,-\frac{3}{2},-\frac{3}{2},\gamma_2,\delta_2,f(\tau)\right]\,e^{h_2(k,\tau)} + i\,C_4\, {\cal H_C}\left[\alpha_2,\frac{3}{2},-\frac{3}{2},\gamma_2,\delta_2,f(\tau)\right]\,e^{-h_2(k,\tau)},
\end{equation}
where $C_3, C_4$ are constants to be determined, and
\begin{eqnarray}
\alpha_2 & = & \frac{k}{H_0 \sqrt{z}}\sqrt{\frac{2\,\left[\theta(z)-1\right]+z }{\left[\theta(z)-1\right]}}, \qquad
\gamma_2 = \frac{k^2\left[2\left[\theta(z)-1\right]+z\right]}{4z\,H^2_0},  \\
\delta_2 &=& \frac{\left[3H^2_0 z-4k^2\right]\theta(z)-3H^2_0 z-2k^2(z-2)}{8 \left[\theta(z)-1\right]z H^2_0}, \\
h_2(k,\tau) & = & \frac{\left[z^{3/2}+2\sqrt{z} \left[\theta(z)-1\right]\right]3H_0\tau(\sqrt{z}-z \left[2\left[\theta(z)-1\right]+z\right])}{2z \left[z+2\left[\theta(z)-1\right]\right]} \nonumber \\
&+& \frac{k \,z^{1/2} e^{-2H_0\tau} \left[\theta(z)-1\right] }{2H_0 \left[z+2\left[\theta(z)-1\right]\right]}\sqrt{\frac{2\left[\theta(z)-1\right]+z}{\left[\theta(z)-1\right]}}.
\end{eqnarray}
The spatial solutions of the modes are the same for both fields [see (\ref{sep})]:
\begin{eqnarray}
R_{kl}(r) &=& \frac{A_1}{\sqrt{r}} \,{\cal LP}\left[a,b, \frac{\sqrt{1+H^2_0 r^2}}{2}\right] +
\frac{B_1}{\sqrt{r}} \,{\cal LQ}\left[a,b, \frac{\sqrt{1+H^2_0 r^2}}{2}\right] , \\
Y_{lm}(\vartheta,\phi) & =& e^{i\,m\,\phi}\,\left[A_2 \,{\cal LP}\left[l,m, \cos(\vartheta)\right] +
B_2 \,{\cal LQ}\left[l,m, \cos(\vartheta)\right]\right],
\end{eqnarray}
where ${\cal LP}\left[a,b, \frac{\sqrt{1+H^2_0 r^2}}{2}\right]$ and ${\cal LQ}\left[a,b, \frac{\sqrt{1+H^2_0 r^2}}{2}\right]$ are respectively the Legendre functions of first and second kind, with parameters
\begin{equation}
a= -\left[\frac{1-\sqrt{1-(k/H_0)^2}}{2}\right], \qquad b=l+1/2,
\end{equation}
such that, for integer $m$-values, we have $-l \leq m \leq l$. Furthermore, we must consider that $k^2>{\cal M}^2 \left[\theta(z)-1\right]\geq 0$ to obtain $\gamma_1>0$, $9H^2_0 > 4 {\cal M}^2 \left[2\left[\theta(z)-1\right]+z\right]$ in order for $h_1$ be real, and finally we must require that $0< k^2 < H_0^2$, in order for $a$ be real.

\subsection{Normalization of the $\hat{\varphi}$-modes with quantum gravity included}

\begin{figure}
    \centering
    \includegraphics[scale=0.6]{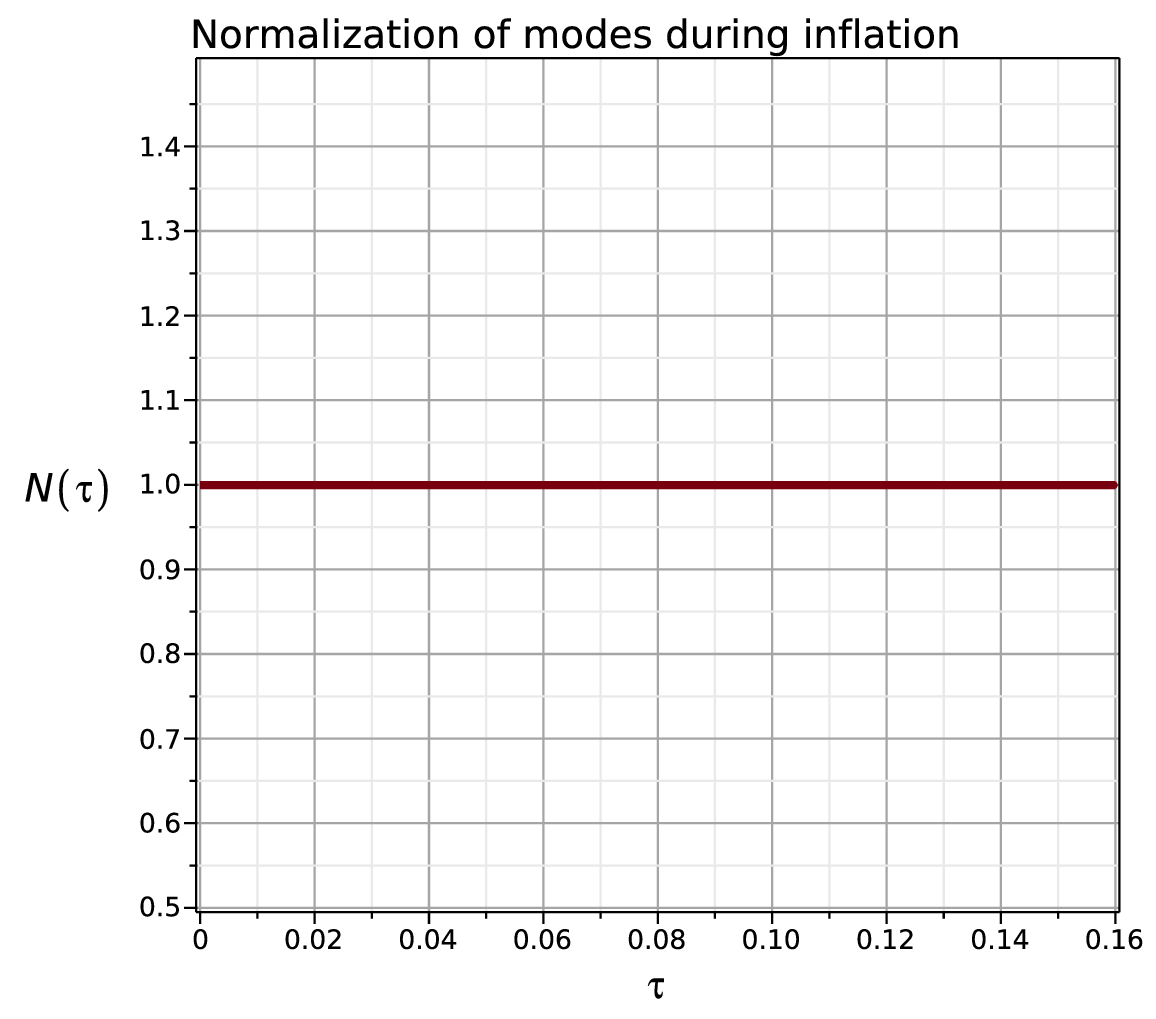}
    \caption{Evolution of $N(\tau)$ during inflation for $z=0.998$ and $0<\tau < 1.6\times 10^{-1}\,{\rm met.}$ Notice that $N(\tau)\simeq 1$ during the early stages of the universe on sub-Hubble scales.}
    \label{f2gi148}
\end{figure}

We must use the equations (\ref{cc1}) and (\ref{cc2}) in the Fourier expansions (\ref{f1}) and (\ref{f2}), in the commutation relationships (\ref{cua1}) and (\ref{c1}), in order for quantize the inflaton field with gravity included. With these expressions we obtain the following normalization condition for the modes:
\begin{equation}\label{nor}
\frac{\partial}{\partial x^{\mu}} \left[{\bf ln}\left(\frac{\varphi_{klm}^*(r,\vartheta,\phi)}{\varphi_{klm}(r,\vartheta,\phi)}\right)\right]=i\,u_{\mu} \frac{\|\Omega_{klm}(r,\vartheta,\phi)\|^2}{\|\varphi_{klm}(r,\vartheta,\phi)\|^2}.
\end{equation}
In our case we are considering a co-moving frame, which is a particular solution of Eqs. (\ref{v0}) and (\ref{v1}), so that $u_0= \sqrt{\bar{g}_{00}}$ and $u_i=0$. Therefore, the normalization condition for the modes (\ref{nor}), results to be
\begin{equation}
\frac{\partial}{\partial x^{0}} \left[{\bf ln}\left(\frac{\xi_k^*(\tau)}{\xi_k(\tau)}\right)\right]=i\,\sqrt{\bar{g}_{00}} \frac{\|\chi_k(\tau)\|^2}{\|\xi_k(\tau)\|^2},
\end{equation}
where $x^0\equiv \tau$ and $\xi_k(\tau)$ and $\chi_k(\tau)$ are the complex functions given respectively by the expressions (\ref{xi}) and (\ref{chi}). Notice that
the normalization of the time dependent inflaton field modes: $\xi_k(\tau)$, are related with the geometric scalar field modes: $\chi_k(\tau)$. This is because the quantization of the physical field is related to the geometry of spacetime. In this framework, we can define the normalization parameter: $N(k,\tau)$:
\begin{equation}
N(k,\tau) = \Im\left[\frac{\partial}{\partial x^{0}} \left[{\bf ln}\left(\frac{\xi_k^*(\tau)}{\xi_k(\tau)}\right)\right]\frac{\|\xi_k(\tau)\|^2}{\sqrt{\bar{g}_{00}}\,\|\chi_k(\tau)\|^2}\right]\simeq 1,
\end{equation}
which must be very close to unity on sub-Hubble scales when the time-dependent inflaton field modes are quantized. In the Fig. (\ref{f2gi148}) we have plotted $N(\tau)\equiv N(k=0.01\,H_0,\tau)$ during inflation, for $C_1=1$, $C_2=10$, $C_3=6.7\times 10^{-14}$ and $C_4=6.7\times 10^{-13}$. Therefore, during the inflationary stage, the modes $\xi_k(\tau)$ remain quantized on sub-Hubble scales with respect to the time-dependent quantum geometric modes $\chi_k(\tau)$.

\subsection{Spacetime fluctuations}

In the following, we will calculate the quadratic spacetime fluctuations during inflation from the expectation value over the Riemann background defined by the metric tensor $\bar{g}_{\alpha\beta}$. In this context, such fluctuations are associated with a zero net flux when varying the action: $\left(\hat{\Omega}^{\alpha}\right)_{; \alpha}=0$, and they have no net effect on Einstein's equations (\ref{eins}), but they will still be responsible for the further structure's formation in the universe:
\begin{equation}
\left.\left<V\right| \hat{\Omega}(\tau, \vec{r}) \,\hat{\Omega}(\tau, \vec{r}')\left|V\right>\right|_{r=r',\phi=\phi',\vartheta=\vartheta'} = \frac{1}{32 \pi} \sum_{l,m}\,\frac{(2l+1)(l-m)!}{(l+m)!} \,\int^{H_0}_{0} \frac{dk}{k}\,k^2 \,\|R_{kl}(r)\|^2\,\|Y_{lm}(\vartheta,\phi)\|^2\,\|\xi_{k}(\tau)\|^2,
\end{equation}
which is related to the power spectrum $\left.{\cal P}_{\Omega}\right|_{(k,l,m)}(\tau)$, of each $(k,l,m)$-mode of the expansion $\left.\left<V\right| \hat{\Omega}(\tau, \vec{r}) \,\hat{\Omega}(\tau, \vec{r}')\left|V\right>\right|_{r=r',\phi=\phi',\vartheta=\vartheta'}=\frac{1}{2\pi} \sum_{m,l} \int^{H_0}_{0} \,\frac{dk}{k}\,\left.{\cal P}_{\Omega}\right|_{(k,l,m)}(\tau)$:
\begin{equation}
\left.{\cal P}_{\Omega}\right|_{(k,l,m)}(\tau) = \frac{A}{16}\,\frac{(2l+1)(l-m)!}{(l+m)!} \,k^2 \,\|R_{kl}(r)\|^2\,\|Y_{lm}(\vartheta,\phi)\|^2\,\|\xi_{k}(\tau)\|^2,
\end{equation}
where $A$ is a constant or normalization. We can evaluate this spectrum outside the horizon, because CMBR comes from distances that, at the moment of emission were near $70$ times the size of the Hubble horizon.
\begin{figure}
    \centering
    \includegraphics[scale=0.6]{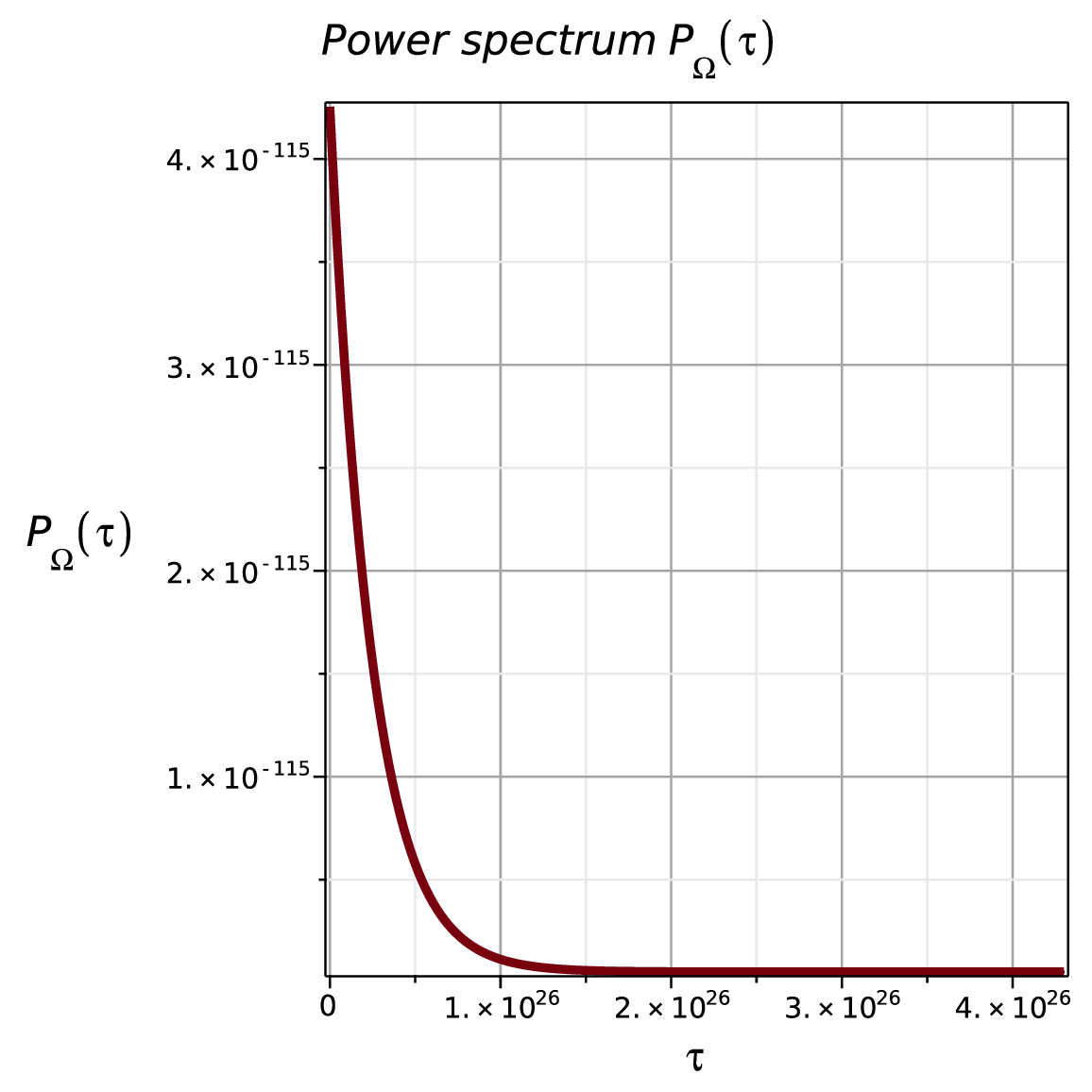}
    \caption{Evolution of ${\cal P}_{\Omega}(\tau)$ after inflation for $8.28\times 10^{23}\,{\rm met.} <\tau < 4.3\times 10^{26}\,{\rm met.}$, $A=1$, $\vartheta=\pi/2$, $l=2$, $m=0$, $k=0.01\,H_0$ and $r=70/H_0$, which were the distances between sources of CMBR that we currently observe. Notice that ${\cal P}_{\Omega}(\tau)$ decreases with time.}
    \label{f3gi148}
\end{figure}
In the Fig. (\ref{f3gi148}) we have plotted the evolution of the spectrum $ {\cal P}_{\Omega}(\tau)\equiv \left.{\cal P}_{\Omega}\right|_{(k=0.01\,H_0,l=2,m=0)}(\tau)$, that comes from the time when the Cosmic Microwave Background Radiation (CMBR) was emitted, that is, when the universe was about $3.8\times 10^{5}$ years old. At that time, the sources from which this radiation was emitted were at distances about 70 times the distance of the Hubble horizon [that corresponds to nearly $8.28\times 10^{23}\,{\rm met.}$. We have plotted this spectrum for times after inflation for $1.18\times 10^{22}\,{\rm met.}<\tau < 4.3\times 10^{26}\,{\rm met.}$, for $A=1$, $\vartheta=\pi/2$, $l=2$, $m=0$, $k=0.01\,H_0$ and $r=70/H_0$, where $r$ is the separation between the sources that emitted the CMBR we observe today. Notice that ${\cal P}_{\Omega}(\tau)\simeq 10^{-115}$ and decreases with time.

\section{Final comments}

We have developed a new general theory of relativity that allows us to study open physical systems on a Riemannian manifold, starting from the new metric tensor $\bar{g}_{\alpha\beta}$, given by $\bar{g}_{\alpha\beta}=g_{\alpha\beta}\left[1+\frac{2\left[\theta(z)-1\right]}{z} \right] -2\,\bar{U}_{\alpha}\bar{U}_{\beta}$ [see Eqs. (\ref{newmetric})]. Therefore the connections $\bar{\Gamma}^{\alpha}_{\hskip 0.15cm \beta\epsilon}$ in (\ref{levi}), are the Levi-Civita connections. The resulting theory is covariant and meets with non-metricity condition that $\bar{g}_{\alpha\beta ; \gamma}=0$, and therefore the same formalization proposed by Einstein in $1915$ is recovered, but for open physical systems where the boundary terms are included in the principle of minimum action. Notice that, due to the complexity of the problem, we have used in the tensor metric components: $\bar{g}_{\alpha\beta}$, a particular solution of the geodesic differential equations for $\bar{U}^{\alpha}$ [see Eq. (\ref{geo})], which is relevant for an observer which is co-moving with the expansion of the universe on cosmological scales. This topic could have been treated more generally by considering the solutions in Eqs. (\ref{v0}) and (\ref{v1}) from the geodesic equations, due to the fact we are considering an universe that is spatially inhomogeneous [we remember that we have considering nonzero spatial curvature]. Our key innovation is the development of a non-perturbative method to quantize simultaneously both, the inflaton field $\hat{\varphi}$ and the fluctuating gravitational field $\hat{\Omega}$. Such innovation incorporate the quantum feedback effects of spacetime on the inflaton field dynamics and the frame of the relativistic observer, which moves in a co-moving reference frame with the tidal wave of dark energy associated with the flow of spacetime fluctuations. It is important to notice that the frame of the relativistic observer is very important in the quantization of the massive inflaton field $\hat{\varphi}$ as one can see in Eqs. (\ref{c1}) and (\ref{c2}). Another important fact, is that the geometric quantum field complies with a non-linear and noncommutative algebra, as can be seen in Eq. (\ref{m56}). In the case of inflation here studied, we have demonstrated that $\hat{\varphi}$ remains quantized during all the inflationary epoch at sub-Hubble scales of the order of $10^{-1}$ met., for $z=0.998$, but can be demonstrated that the field remains quantized to scales of the order of $10^{26}$ met, which is the size of the present day observable universe. Finally, the spectrum for the quadratic fluctuations of the geometric field: $\Omega$ was calculated and plotted in the Fig. (\ref{f3gi148}) for the particular case where $k=0.01\,H_0$, $l=2$ and $m=0$. The amplitude is really small: ${\cal P}_{\Omega}(\tau)\simeq 10^{-115}$, and decreases with time. This is due to the fact the gravitational field is very weak.

To conclude, the formalism developed here has multiple applications and could be applied to all types of open relativistic systems in the astrophysical and cosmological fields, thus opening up the scope of the theory of general relativity. Furthermore, it has been shown that the quantization of any physical field associated with a massive field cannot be performed independently of $\Omega$, since said field describes the geometric distortion caused by the physical field.

\section*{CRediT authorship contribution statement}

Alan Sebasti\'an Morales: Ideas, Conceptualization, Calculation, Writing. Mauricio Bellini: Ideas, Conceptualization, Calculation, Writing. Juan Ignacio Musmarra:
Ideas, Conceptualization, Calculation, Writing.

\section*{Declaration of competing interest}

The authors declare the following financial interests/personal relationships which may be considered as potential competing interests:
The authors report financial support provided by both, National Scientific and Technical Research Council (CONICET) and National University of Mar del Plata (UNMdP).

\section*{Acknowledgements}

The author acknowledges CONICET, Argentina (PIP 11220200100110CO), and UNMdP (EXA1156/24) for financial support.

\section*{Data availability}

No data was used for the research described in the article.\\


\bigskip
\begin{thebibliography}{99}
\bibitem{in0} A. Starobinsky, Phys. Lett. {\bf B91}: 99 (1980).
\bibitem{in1} A. Guth, Phys. Rev. {\bf D23}: 347 (1981).
\bibitem{in11} A. A. Starobinsky, Phys. Lett. {\bf B117}: 175 (1982).
\bibitem{in2} A. D. Linde, Phys. Lett. {\bf B108}: 389 (1982).
\bibitem{in3} A. Albrecht and P. J. Steinhardt, Phys. Rev. Lett. {\bf 48}: 1220 (1982).
\bibitem{in33} S. W. Hawking, I. G. Moss, J. M. Stewart, Phys. Rev. {\bf D26}: 2681 (1982).
\bibitem{in333} S. W. Hawking and I. G. Moss, Phys. Lett. {\bf B110}: 35 (1982).
\bibitem{hm} S. W. Hawking, I. G. Moss, Nucl. Phys. {\bf B224}: 180 (1983).
\bibitem{in4} A. D. Linde, Phys. Lett. {\bf B129}: 177 (1983).
\bibitem{b1} H. J. de Vega and N. Sánchez, Phys. Rev. {\bf D50}: 7202 (1994).
\bibitem{bcms} M. Bellini, H. Casini, R. Montemayor, P. Sisterna, Phys. Rev. {\bf D54}: 7172 (1996).
\bibitem{vega1} D. Boyanovsky, D. Cormier, H. J. de Vega, R. Holman, S. P. Kumar, Phys. Rev. {\bf D57}: 2166 (1998).
\bibitem{maube1} M. Bellini, Phys. Lett. {\bf B428}: 31 (1998).
\bibitem{maube2} M. Bellini, Nucl. Phys. {\bf B604}: 441 (2001).
\bibitem{vega2} F. J. Cao, H. J. de Vega, N. G. S\'anchez, Phys. Rev. {\bf D70}: 083528 (2004).
\bibitem{vega3} D. Boyanovsky, H. J. de Vega, N. G. S\'anchez, Phys. Rev. {\bf D71}: 023509 (2005).
\bibitem{vega4} H. J. de Vega, N. G. S\,anchez, Phys. Rev. {\bf D74}: 063519 (2006).
\bibitem{mrv} J. Martin, C. Ringeval, V. Vennin, {\it Enciclopaedia Inflationary}. Phys. Dark Univ. {\bf 42}: 101653 (2024).
\bibitem{Kundu2011} S. Kundu, JCAP {\bf 2}: 5 (2012).
\bibitem{Salvio2017} A. Salvio, Phys. Lett. {\bf B780}: 111-117 (2018).
\bibitem{Cai2019} Y. Cai, Y. Piao, Sci. China Phys. Mech. Astron. {\bf 63}: 11, 110411 (2020).
\bibitem{Melcher2023} B. Melcher, A. Pradhan, S. Watson, Phys. Rev. {\bf D110}: 6, 063517 (2024).
\bibitem{Joana2024} C. Joana, Phys. Rev. {\bf D110}: 6, 063534 (2024).
\bibitem{Cognola2009} G. Cognola, E. Elizalde, S. D. Odintsov, P. Tretyakov, S. Zerbini, Phys. Rev. {\bf D79}: 044001 (2009).
\bibitem{Odintsov2021} S. D. Odintsov, V. K. Oikonomou, Phys. Lett {\bf B824}: 136817 (2022).
\bibitem{Powell2006} B. Powell, W. H. Kinney, Phys. Rev. {\bf D76}: 063512 (2007).
\bibitem{Wang2007} I. Wang, K. Ng, Phys. Rev. {\bf D77}: 083501 (2008).
\bibitem{Destri2009} C. Destri, H. J. de Vega, N. G. Sanchez, Phys. Rev. {\bf D81}: 063520 (2010).
\bibitem{Dudas2012} E. Dudas, N. Kitazawa, S. P. Patil, A. Sagnotti, JCAP {\bf 5}: 12 (2012).
\bibitem{Das2014} S. Das, G. Goswami, J. Prasad, R. Rangarajan, JCAP {\bf 6}: 6, 1 (2015).
\bibitem{Kitazawa2014} N. Kitazawa, A. Sagnotti, JCAP {\bf 4}: 17 (2014).
\bibitem{Gruppuso2016} A. Gruppuso, N. Kitazawa, N. Mandolesi, P. Natoli, A. Sagnotti, Phys. Dark Univ. {\bf 11}: 68-73 (2016).
\bibitem{Jin2018} W. Jin, Y. Ma, T. Zao, JCAP {\bf 2}: 10 (2019).
\bibitem{Agullo2013} I. Agullo, A. Ashtekar, W. Nelson, Class. Quant. Grav. {\bf 30}: 085014 (2013).
\bibitem{Ashtekar2015} A. Ashtekar, A. Barrau, Class. Quant. Grav. {\bf 32}: 23, 234001 (2015).
\bibitem{Zhu2017} T. Zhu, A. Wang, G. Cleaver, K. Kirsten, Q. Sheng, Phys. Rev. {\bf D96}: 8, 083520 (2017).
\bibitem{Mohammadi2024} A. Mohammadi, JCAP {\bf 10}: 62 (2024).
\bibitem{Shahalam2025} M. Shahalam, Int. J. Geom. Meth. Mod. Phys. {\bf 22}: 7, 2550030 (2025).
\bibitem{np} M. Bellini. Nucl. Phys. {\bf B563}: 245 (1999).
\bibitem{boj} M. Bojowald. Liv. Rev. Rel. {\bf 8}: 11  (2005).
\bibitem{cher} S. Cheraghchi, F. Shojai, M. H. Abbassi. Phys. Scripta {\bf 98}: 085310 (2023).
\bibitem{bit} M. Bitaj, N. Rashidi, K. Nozari, M. Roushan. Phys. Dark Univ. {\bf 46}: 101552 (2024).
\bibitem{day} A. C. Day, I. A. Brown. {\bf JCAP 03}: 005 (2014).
\bibitem{el} E. Abdalla, G. F. Abell\'an, A. Aboubrahim, A. Agnello, O. Akarsu {\em et al.}. JHEAp {\bf 34}: 49 (2022).
\bibitem{Bellini} M. Bellini, Eur. Phys. J. {\bf C84}: 1270 (2024).
\bibitem{York} J. W. York, Phys. Rev. Lett. {\bf 16}: 1082 (1972).
\bibitem{Gibbons} G. W. Gibbons, S. W. Hawking, Phys. Rev. {\bf D15}: 2752 (1977).
\bibitem{Parattu} K. Parattu, S. Chakraborty, B. R. Majhi, T. Padmanabhan, Gen. Rel. Grav. {\bf 48}: 94 (2016).
\bibitem{rb1} L. S. Ridao, M. Bellini, Astrophys. Space Sci. {\bf 357}: 94 (2015).
\bibitem{univ} M. Bellini, J. I. Musmarra, P. A. S\'anchez, A. S. Morales. Universe {\bf 11}: 243 (2025).
\bibitem{ricci1} M. Rumpf. Phys. Rev. {\bf D33}: 942 (1986).
\bibitem{ricci-flow} M. Bellini. Phys. Scripta {\bf 96}: 065301 (2021).
\bibitem{pal} A. Palatini. {\it Deduzione invariantiva delle equazioni gravitazionali dal principio di Hamilton}, Rend. Circ. Mat. Palermo {\bf 43}: 203-212  (1919). [English translation by R.Hojman and C. Mukku in P.G. Bergmann and V. De Sabbata (eds.) Cosmology and Gravitation, Plenum Press, New York (1980)].
\bibitem{p2} M. Bellini. Ann. Phys. {\bf 476}: 169973 (2025).
\bibitem{BMAS} M. Bellini, J. I. Musmarra, M. Anabitarte, P. A. S\'anchez. Phys. Dark Univ. {\bf 42}: 101359 (2023).
\bibitem{mabe} D. Magos Cortes, M. Bellini, Astrop. Phys. {\bf 163}: 103006 (2024).
\end{thebibliography}
\end{document}